\documentclass[english]{article}
\usepackage[utf8]{inputenc}
\usepackage[T1]{fontenc}
\usepackage{babel}
\usepackage{amsmath}
\usepackage{graphicx}
\usepackage{subcaption}
\usepackage{tikz}
\usepackage{hyperref}
\hypersetup{colorlinks=true, breaklinks=true, linkcolor=blue, urlcolor=blue, citecolor=blue}

\begin{document}

\title{A Cone-Beam X-Ray CT Data Collection designed for Machine Learning}

\author{Henri Der Sarkissian\textsuperscript{(1,{*})}, Felix Lucka\textsuperscript{(1,2,{*})}, Maureen van Eijnatten\textsuperscript{(1)}, \\
Giulia Colacicco\textsuperscript{(1)}, Sophia Bethany Coban\textsuperscript{(1)}, Kees Joost Batenburg\textsuperscript{(1,3)}}

\maketitle
\noindent (1) Centrum Wiskunde en Informatica, Computational Imaging group, Science Park 123, 1098XG Amsterdam, The Netherlands \\
(2) Centre for Medical Image Computing, University College London, WC1E 6BT London, United Kingdom \\
(3) Leiden University, Department of Mathematics, 2300 RA Leiden, The Netherlands \\
{(*)} Corresponding authors:
Henri Der Sarkissian (henri.dersarkissian@gmail.com) \& Felix Lucka (Felix.Lucka@cwi.nl)

\begin{abstract}
\noindent
Unlike previous works, this open data collection consists of X-ray cone-beam (CB) computed tomography (CT) datasets specifically designed for machine learning applications and high cone-angle artefact reduction. Forty-two walnuts were scanned with a laboratory X-ray set-up to provide not only data from a single object but from a class of objects with natural variability. For each walnut, CB projections on three different source orbits were acquired to provide CB data with different cone angles as well as being able to compute artefact-free, high-quality ground truth images from the combined data that can be used for supervised learning. We provide the complete image reconstruction pipeline: raw projection data, a description of the scanning geometry, pre-processing and reconstruction scripts using open software, and the reconstructed volumes. Due to this, the dataset can not only be used for high cone-angle artefact reduction but also for algorithm development and evaluation for other tasks, such as image reconstruction from limited or sparse-angle (low-dose) scanning, super resolution, or segmentation.
\end{abstract}

\clearpage

\section{Background \& Summary}

\subsection{Scientific Context}

\begin{figure}[bh!]
    \centering
    \includegraphics[]{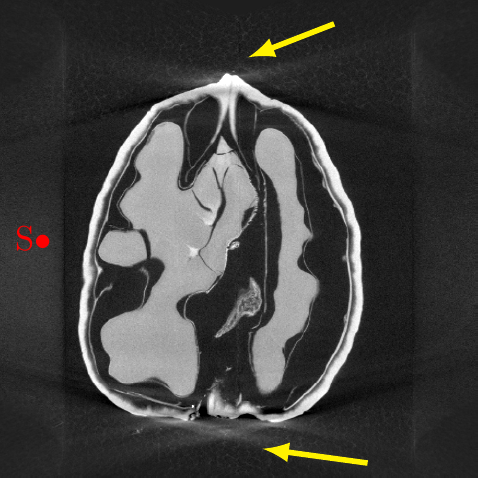}
    
    \caption{Vertical slice through an FDK reconstruction of a CBCT scan of a walnut. The red dot indicates the vertical position of the source orbit and the yellow arrows point at the high cone angle artefacts.}
    \label{fig.cbartefactexample}
\end{figure}

X-ray computed tomography (CT) is a widely used projection-based imaging modality with a broad range of clinical, scientific and industrial applications. In many of those, CT scanners use a particular projection geometry called circular \textit{cone-beam} (\textit{CB}). This scanning geometry typically leads to a distinct type of artefact in the image regions with a high cone angle, cf. Figure~\ref{fig.cbartefactexample}. While several reconstruction or correction methods have been proposed to reduce high cone angle artefacts~\cite{hsieh2000,dennerlein2008,zhang2016}, they remain a crucial drawback of CBCT scanners over other scanners, which have disadvantages such as higher radiation dose or costs in return~\cite{KoEiKiShWo17}.

In the scientific field, there is often a clear division between computational imaging groups with a background in mathematics and computer science, which focus on enhancing CT methodology on one side and experimental imaging groups using CT as a tool to conduct their scientific studies on the other. The latter typically use commercial CT solutions coming with proprietary software which does not give full access to the raw projection data or the details of the experimental acquisition. As a result, many mathematical and computational studies rely on artificial data simulated with varying degrees of realism. This lack of suitable experimental data is a significant hurdle for the translation of innovative research into applications. 

Many important, recent CT innovations introduce machine learning techniques into the tomographic image reconstruction process~\cite{wang2018,ravishankar2019}, in particular deep neuronal networks (\textit{deep learning}). For these approaches, realistic experimental data is not only needed for evaluation but more crucially, for constructing the method itself. Namely, the network parameters are optimized based on \textit{training data} which consists of a large number of representative pairs of input data with the desired ideal output of the network (\textit{ground truth}). While many large, open, bench-mark data collections meeting these criteria exist for standard applications of deep learning (e.g., MNIST \cite{MNIST} for the classification of handwritten digits), there are very few suitable projection datasets for deep learning for CT so far.

\subsection{Previous Works \& Limitations}

Several open fan beam (2D) and cone beam (3D) X-ray CT datasets acquired by a laboratory set-up or with synchrotron parallel X-ray sources exist~\cite{coban2015,jorgensen2017,singh2018,decarlo2018,walnuthelsinki}. Suitable clinical datasets are more difficult to acquire and distribute openly. The ultimate quality measure for clinical images is their diagnostic value, which needs to be assessed by radiologists. Therefore, data is often only distributed as part of an image reconstruction challenge. A prominent example of this is the Mayo Clinic Low Dose CT challenge~\cite{lowdosechallenge} consisting of 3D helical CT abdominal scans of ten cancer patients. While these datasets are extremely useful to evaluate reconstruction algorithms on a wide range of different objects and acquisition conditions, they are not suitable for machine learning as they typically contain only a single or very few scanned objects or have not been designed such that the reconstruction quality can easily be assessed in an automated way with respect to a high-quality ground truth reconstruction. 

\subsection{Motivation \& Summary}

To fill this gap, we acquired a carefully designed CBCT data collection suitable for developing machine learning approaches: 42 walnuts (this choice will be discussed in the next section) were scanned with a special laboratory X-ray CBCT scanner. For each sample, CB projections were acquired on 3 different circular orbit heights. This creates different cone angles and resulting artefact pattern as well as allowing for an artefact-free, high-quality ground truth to be computed from the combined data. We provide reconstructed volumes and an open software implementation of the complete image reconstruction pipeline along with the raw projection data. Note that while 42 samples seem few compared the training data sizes used in other deep learning applications, each sample here is a 3D object. Extracting 2D slices from these high-resolution volumes composed of $501^3$ voxels gives enough data for training 2D networks that are then used to process volumes slice-by-slice, which is currently the most common approach in 3D applications~\cite{PeBaSe18}. While this dataset is designed to benchmark machine-learning-based correction techniques for CB artefacts, it can also be used for algorithm development and evaluation for other tomography applications, such as image reconstruction from limited or sparse-angle (low-dose) scanning, super resolution, or for image analysis tasks such as semantic segmentation.


\section{Methods} \label{sec:Methods}

\subsection{Sample Collection}

A data set suitable for deep learning with convolutional layers (\textit{convolutional neuronal networks, CNNs}) needs to be collected in a particular way. During training, the network needs to learn to recognize common spatial features and their natural variations of the class of objects that should be imaged. For this, data acquired from a sufficiently large number of representative samples is needed. Having too few samples to train on can lead to over-fitting and reduce the network's ability to generalize to unseen data. Partly inspired by~\cite{walnuthelsinki}, we decided to scan 42 walnuts: Similar to objects scanned in (pre-)clinical imaging, they contain natural inter-population variability which is advantageous compared to manufactured objects like phantoms used to calibrate scanners. They consist of a hard shell, a softer inside, air filled cavities and a variety of large-to-fine-scale details which makes them a good proxy for the human head. In addition, their size ($\approx 3~\mathrm{cm}$ height) is suitable for our experimental set-up.    

\subsection{X-Ray Tomography Scanner}

\begin{figure}
    \centering
    \includegraphics[]{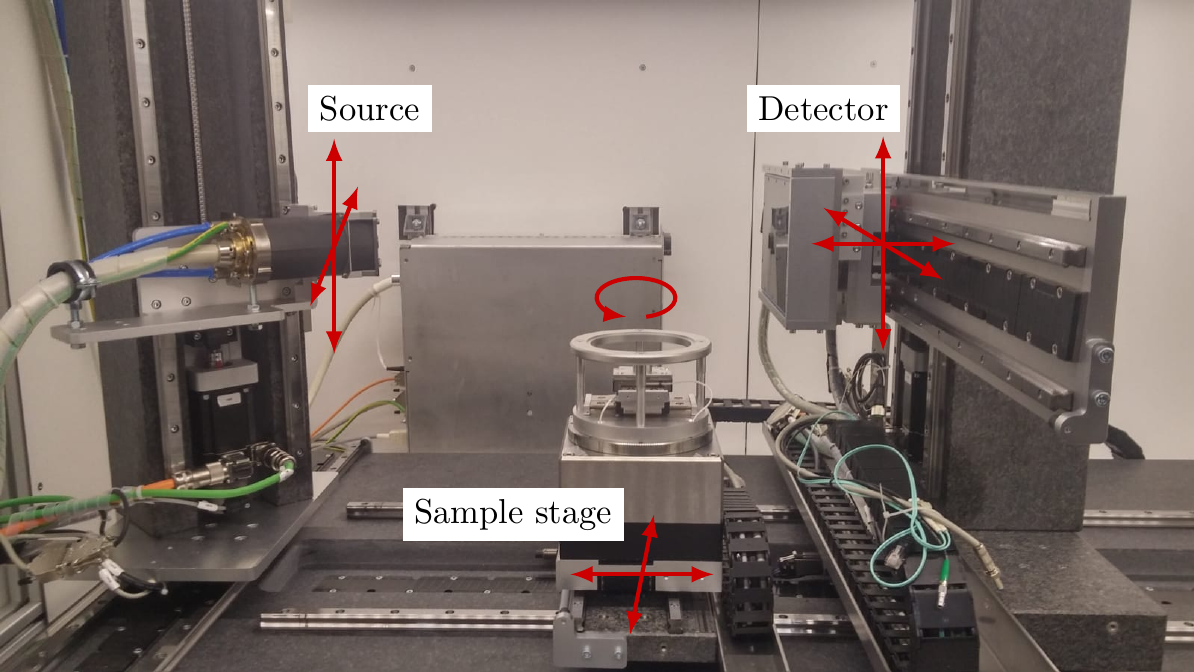}
    \caption{ FleX-ray Lab, the X-ray cone-beam tomography set-up used for the data acquisition~\cite{flexray}. The arrows indicate the degrees of freedom.}
    \label{fig.flexray}
\end{figure}

The scans were performed using a custom-built, highly flexible X-ray CT scanner, the FleX-ray scanner, developed by XRE nv\footnote{\href{https://xre.be/}{https://xre.be/}} and located in the FleX-ray Lab at the Centrum Wiskunde \& Informatica (CWI) in Amsterdam, Netherlands~\cite{flexray}. The general purpose of the FleX-ray Lab is to conduct proof-of-concept experiments directly accessible to researchers in the field of mathematics and computer science. 
The scanner consists of a cone-beam microfocus X-ray point source (limited to 90 kV and 90 W) that projects polychromatic X-rays onto a $1536\times 1944$ pixels, $74.8~\mathrm{\mu m}^2$ each, 14-bit flat panel detector (Dexella 1512NDT) and a rotation stage in-between, upon which the sample is mounted. All three components are mounted on translation stages which allow them to move independently from one another. A schematic view of the set-up with the description of possible movements is shown in Figure~\ref{fig.flexray}. 

\subsection{Projection Geometry and Acquisition Parameters}

\begin{figure}
    \centering
        \includegraphics[height=10cm]{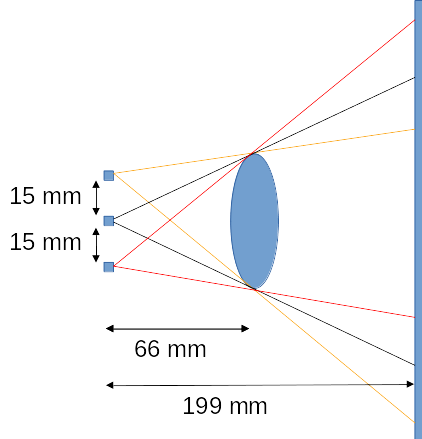} \\
        \vspace{0.3cm}
        \includegraphics[height=3.9cm]{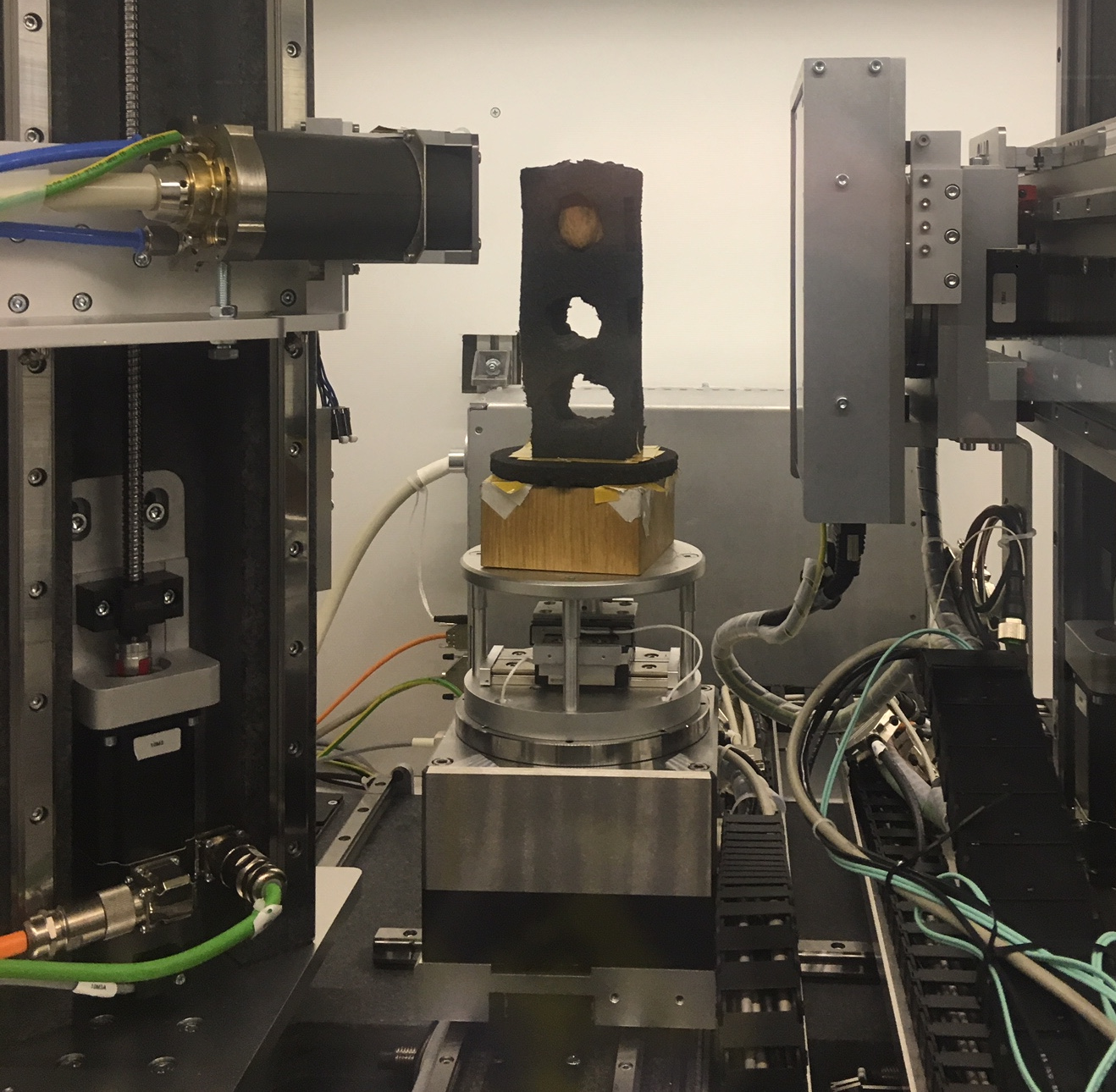}
        \includegraphics[height=3.9cm]{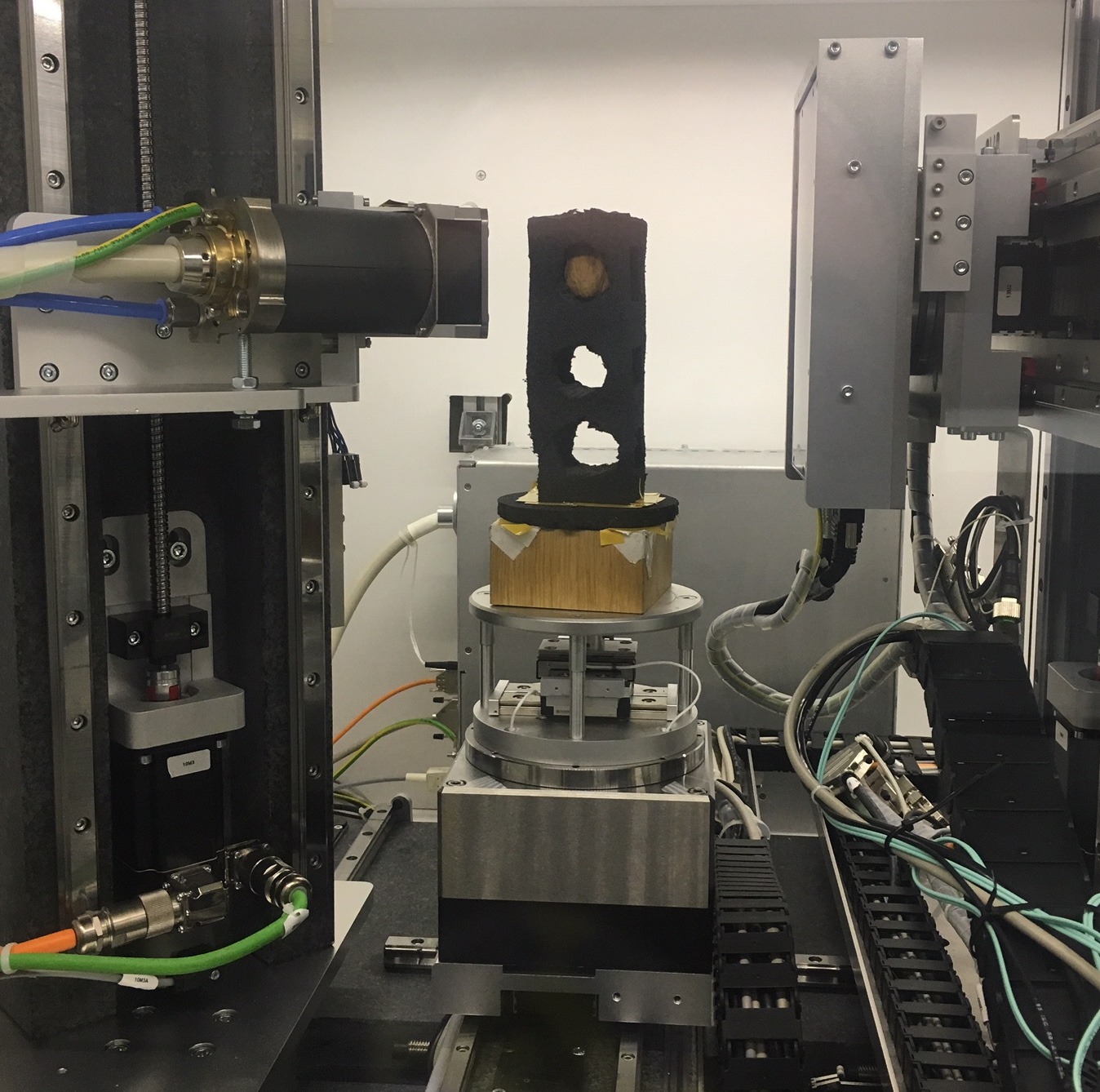}
        \includegraphics[height=3.9cm]{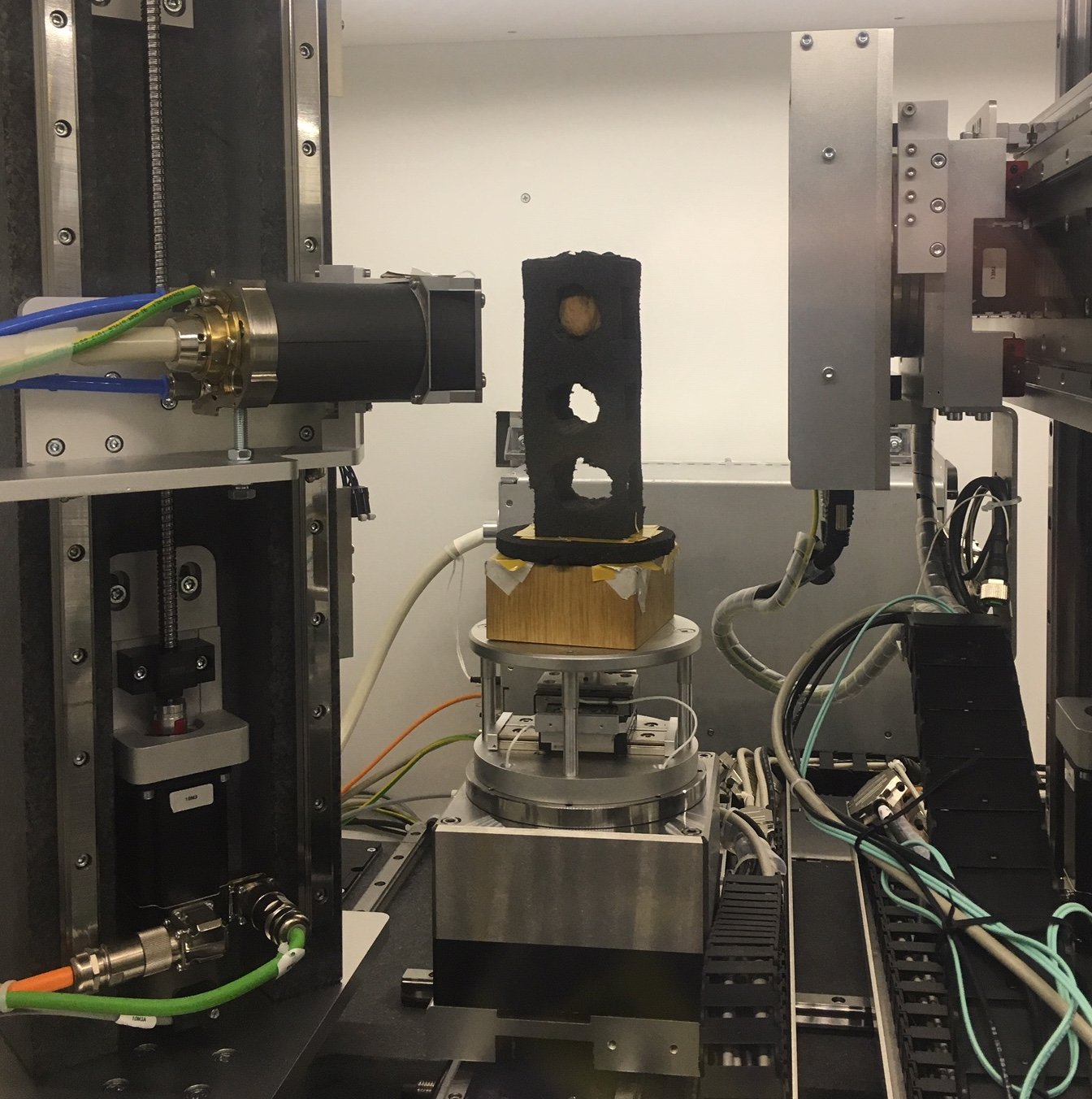}
    \caption{Scanning geometry and trajectories for each sample. Top row: Schematic view from the side. Three full circular orbits are recorded at 3 distinct source and detector heights. The 3 squares on the left denote the source positions. Bottom row: Photographs of actual realization.}
    \label{fig.trajectories}
\end{figure}

Our aim was to create a data collection suitable for \textit{supervised learning}. In supervised learning, the training data consists of pairs of input data with the desired ideal output of the network (the ground truth). A distance function (\textit{training loss}) between ground truth and current output of the network is used to drive the optimization of the network's parameters. In our case, the input to the network may be the artefact-ridden reconstruction of a sample computed from a single orbit CBCT data set, and the ground truth could be the corresponding, high-quality, artefact-free reconstruction. We thus needed to acquire projection data from which both of these reconstructions can be computed. To obtain severe high cone angle artefacts, we needed to maximize the vertical cone-beam angle by moving the sample as close as possible to the source and choose an appropriate detector-to-object distance to maximize magnification while keeping the sample in the field of view at all times. Then, we varied the source height to collect projections from 3 circular orbits, cf. Figure~\ref{fig.trajectories} (the detector height needed to be adjusted accordingly in order to fit the entire sample in every projection). In the following section, we will see that while the reconstructions from each orbit alone have different artefact pattern, combining the data from all orbits gives a high-quality reconstruction free of high cone angle artefacts.

Each walnut was embedded in a foam mount (cf. Figure~\ref{fig.trajectories} bottom row). This foam is almost transparent to the X-ray beam used in our experiment. For each orbit, 1201 projections were taken during a continuous, full rotation of the sample. First and last projection were taken at the same position, leading to an angular increment of $0.3^\circ$. The exposure time for each projection was $80 \mathrm{ms}$ and the acquired data was binned on the fly by 2-by-2 pixel windows, i.e. each raw projection was of size $768\times972$ pixels. Each binned detector pixel is sized $149.6\times 149.6~\mu\mathrm{m}^2$ for a total detection field of view of $114.89\times 145.41~\mathrm{mm}^2$. During the experiment, the source voltage and power were set to $40~\mathrm{kV}$ and $12~\mathrm{W}$, respectively. These values had been adjusted to ensure maximum contrast in the projection domain while avoiding detector saturation. Table~\ref{tab.scannerconfig} summarizes the acquisition parameters used.

Before every orbital scan, the source was turned off to record a projection of the detector offset count, the so-called \textit{dark-field} image. After switching the source on again, a projection was recorded without the sample in the field of view, the so-called \textit{flat-field} image showing the beam profile. A second flat-field was recorded after the orbital scan to correct for shadowing effects. Flat-field and dark-field images can be used to pre-process the raw photon count data for the image reconstruction as described in the next section. Examples of the projections collected for each sample are shown in Figure~\ref{fig.projex}. 

\begin{table}
\centering
\begin{tabular}{ll}
    \hline
    Tube voltage & $40~\mathrm{kV}$ \\
    Tube power & $12~\mathrm{W}$ \\
    Exposure time & $80~\mathrm{ms}$ \\
    Number of averages & 1\\
    Hardware binning & $2\times 2$ pixels\\
    Effective detector pixel size & $149.6~\mathrm{\mu m}$\\
    Detector rows & 972\\
    Detector colums & 768\\
    Source to object distance & $66~\mathrm{mm}$\\
    Source to detector distance & $190~\mathrm{mm}$ \\
    Magnification & $3.016$\\
    Number of projections per orbit & $1201$\\
    Angular increment & $0.3^\circ$\\
    \hline
\end{tabular}
\caption{\label{tab.scannerconfig}Summary of the acquisition parameters used.}
\end{table}

\newcommand{\imw}{0.27}
\begin{figure}
    \centering
        \begin{subfigure}[b]{1\textwidth}
            \begin{subfigure}[h]{0.05\textwidth}
                \caption*{}
            \end{subfigure}
            \begin{subfigure}[h]{\imw\textwidth}
                \subcaption*{high source position}
            \end{subfigure}
            \begin{subfigure}[h]{\imw\textwidth}
                \subcaption*{mid. source position}
            \end{subfigure}
            \begin{subfigure}[h]{\imw\textwidth}
                \subcaption*{low source position}
            \end{subfigure}
        \end{subfigure}
        \begin{subfigure}[b]{1\textwidth}
            \begin{subfigure}[c]{0.05\textwidth}
                \caption*{\rotatebox{90}{flat-field}}
            \end{subfigure}
            \begin{subfigure}[c]{\imw\textwidth}
                \includegraphics[angle=90,width=\textwidth]{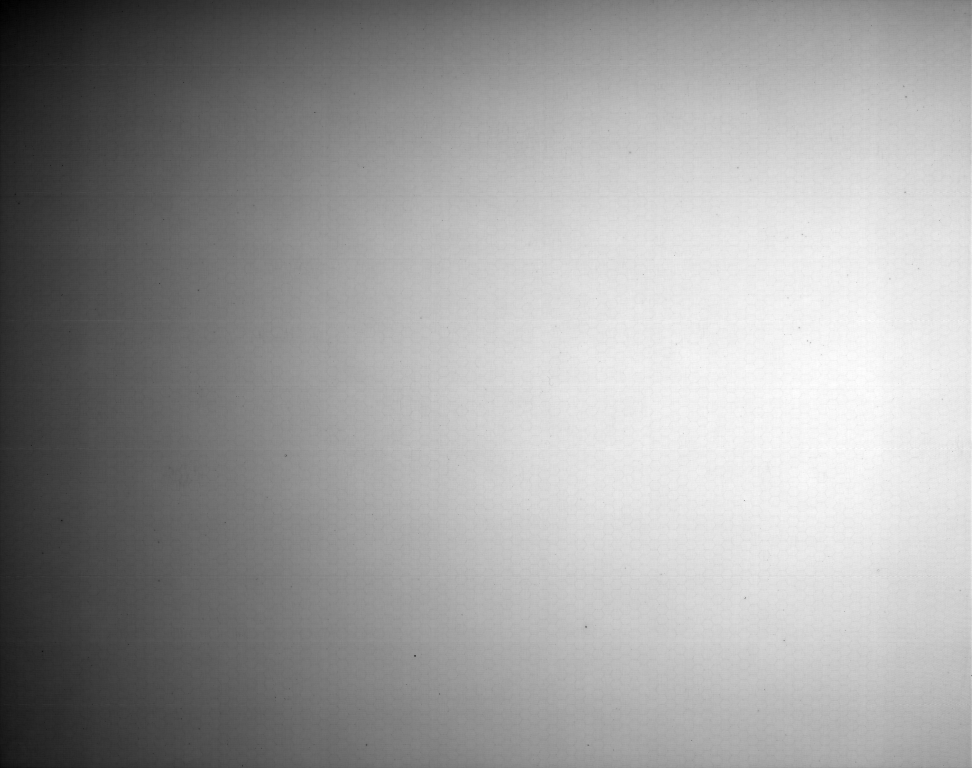}
                \subcaption*{[5899,13342]}
            \end{subfigure}
            \begin{subfigure}[c]{\imw\textwidth}
                \includegraphics[angle=90,width=\textwidth]{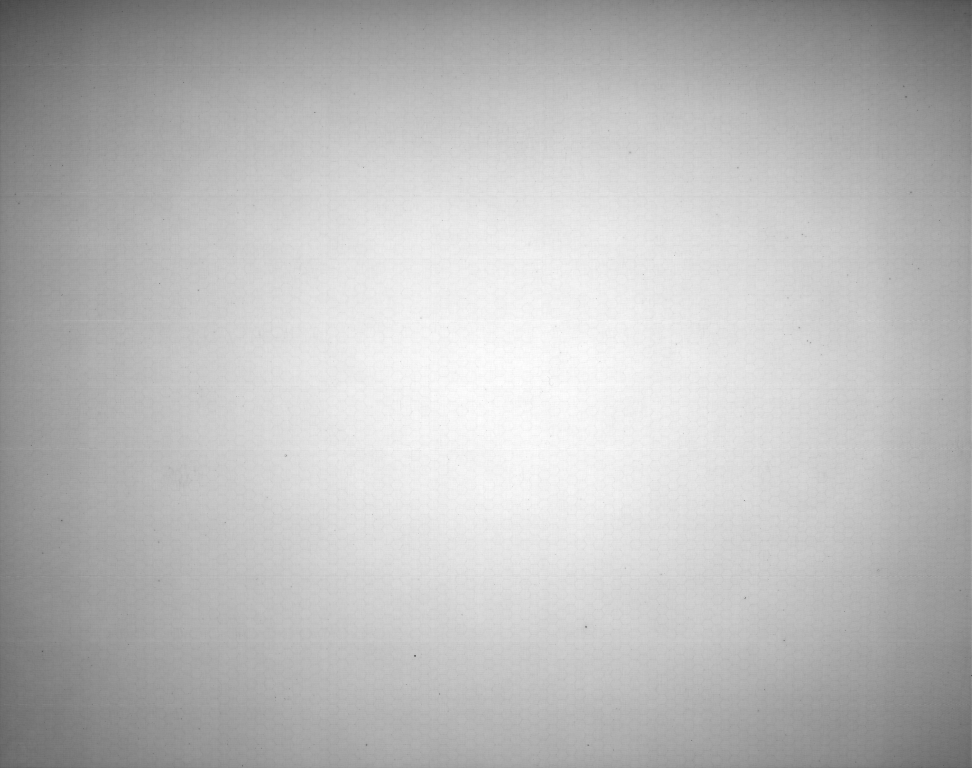}
                \subcaption*{[7983,13314]}
            \end{subfigure}
            \begin{subfigure}[c]{\imw\textwidth}
                \includegraphics[angle=90,width=\textwidth]{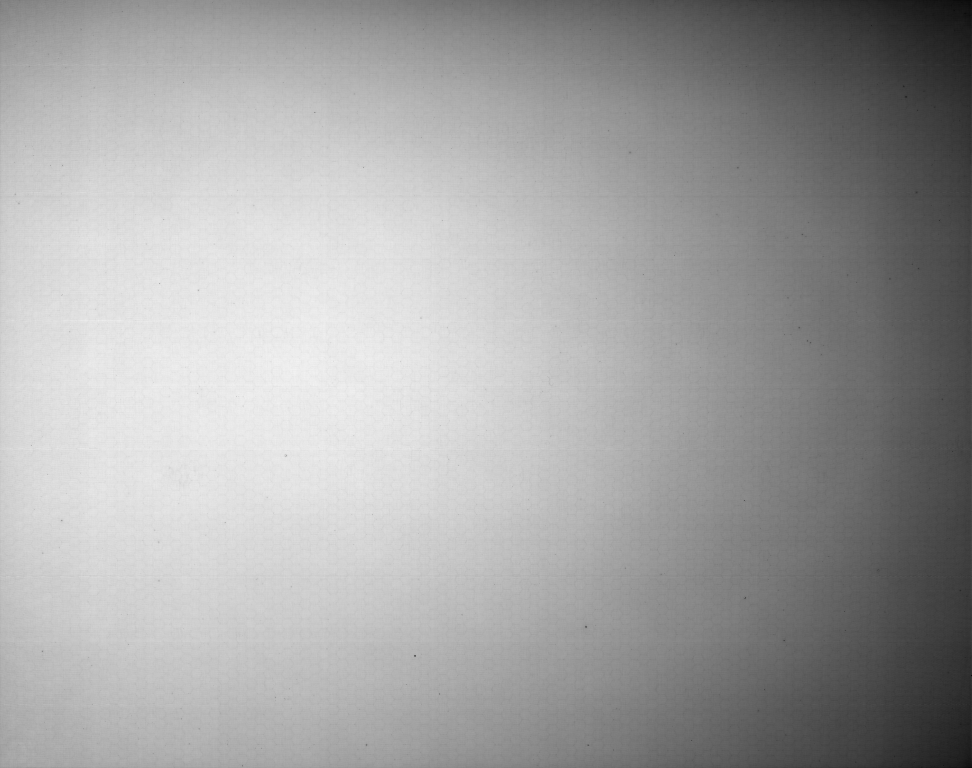}
                \subcaption*{[6281,12970]}
            \end{subfigure}
        \end{subfigure}
        \begin{subfigure}[b]{\textwidth}
            \begin{subfigure}[c]{0.05\textwidth}
                \caption*{\rotatebox{90}{raw projections}}
            \end{subfigure}
            \begin{subfigure}[c]{\imw\textwidth}
                \includegraphics[angle=90,width=\textwidth]{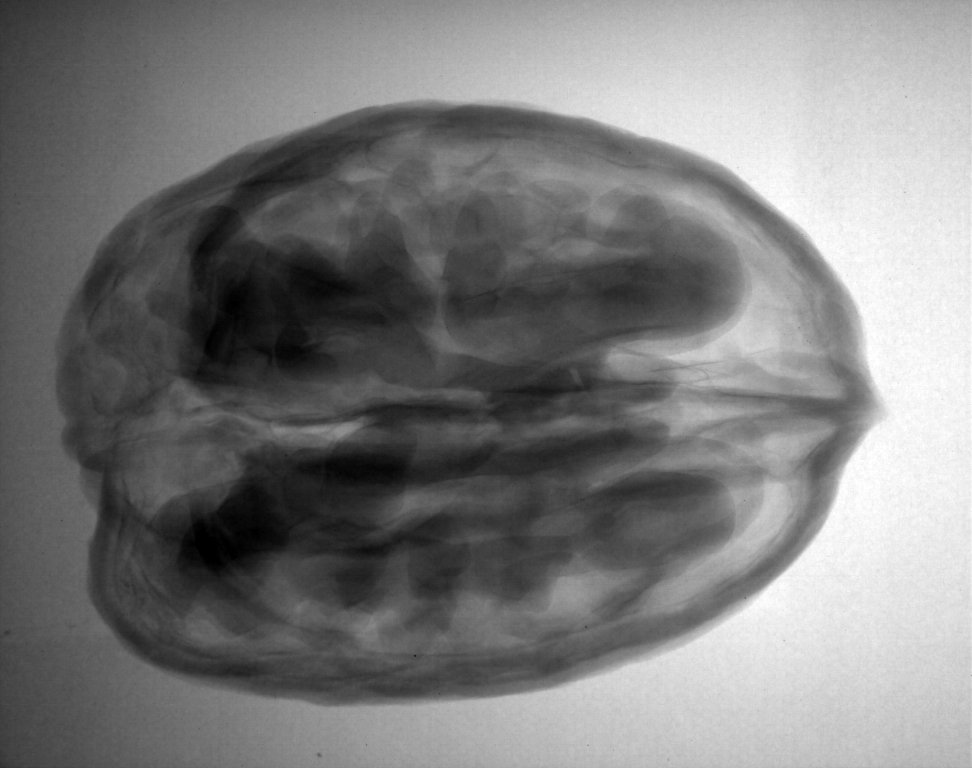}
                \subcaption*{[3975,12734]}
            \end{subfigure}
            \begin{subfigure}[c]{\imw\textwidth}
                \includegraphics[angle=90,width=\textwidth]{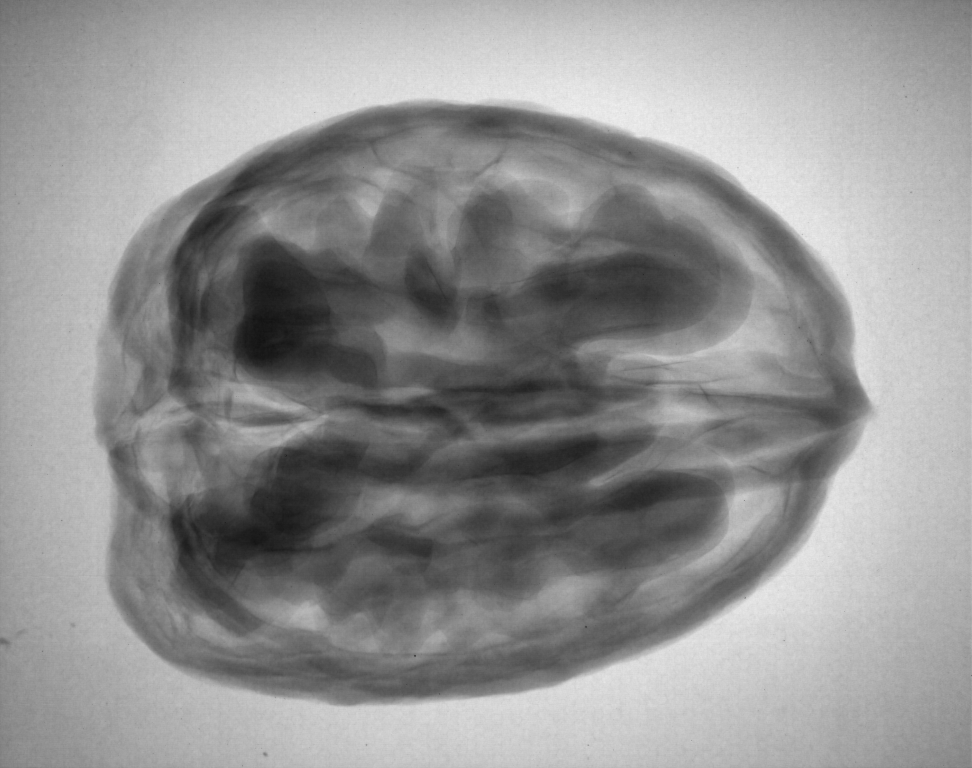}
                \subcaption*{[4918,12077]}
            \end{subfigure}
            \begin{subfigure}[c]{\imw\textwidth}
                \includegraphics[angle=90,width=\textwidth]{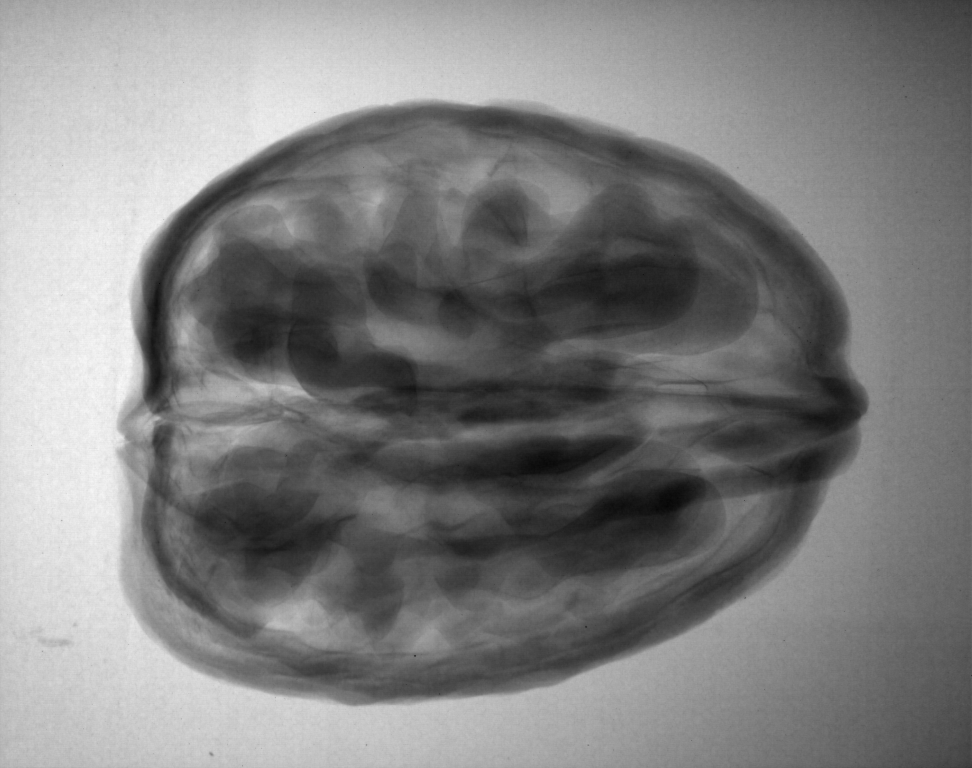}
                \subcaption*{[4544,12340]}
            \end{subfigure}
        \end{subfigure}
        \begin{subfigure}[b]{\textwidth}
            \begin{subfigure}[c]{0.05\textwidth}
                \caption*{\rotatebox{90}{dark-field}}
            \end{subfigure}
            \begin{subfigure}[c]{\imw\textwidth}
                \includegraphics[angle=90,width=\textwidth]{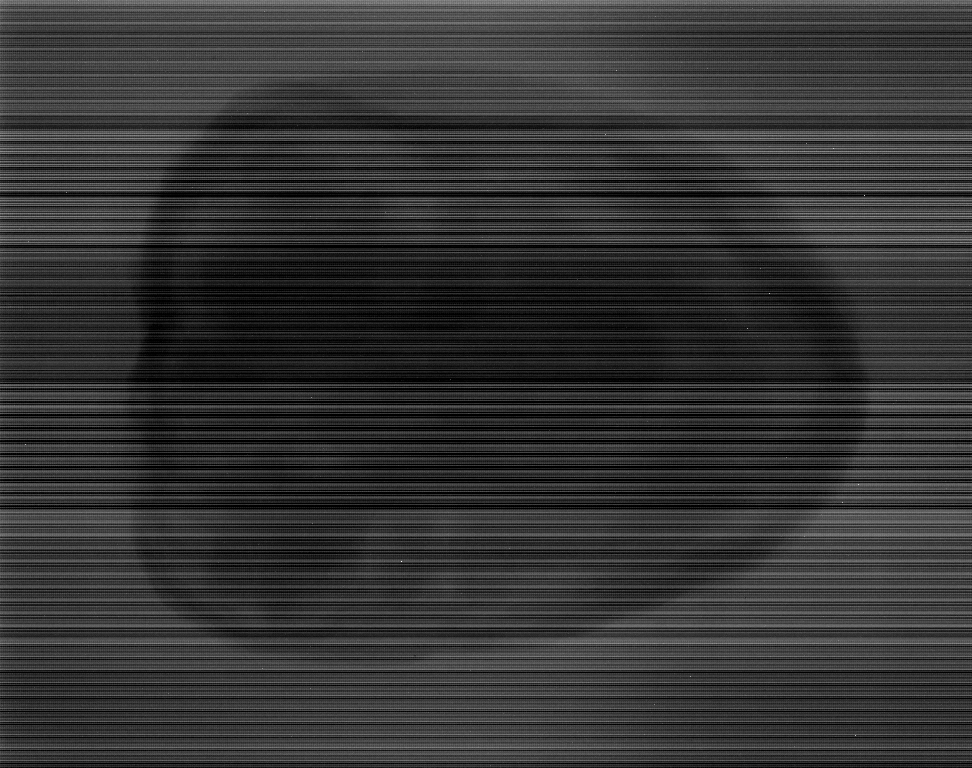}
                \subcaption*{[304,410]}
            \end{subfigure}
            \begin{subfigure}[c]{\imw\textwidth}
                \includegraphics[angle=90,width=\textwidth]{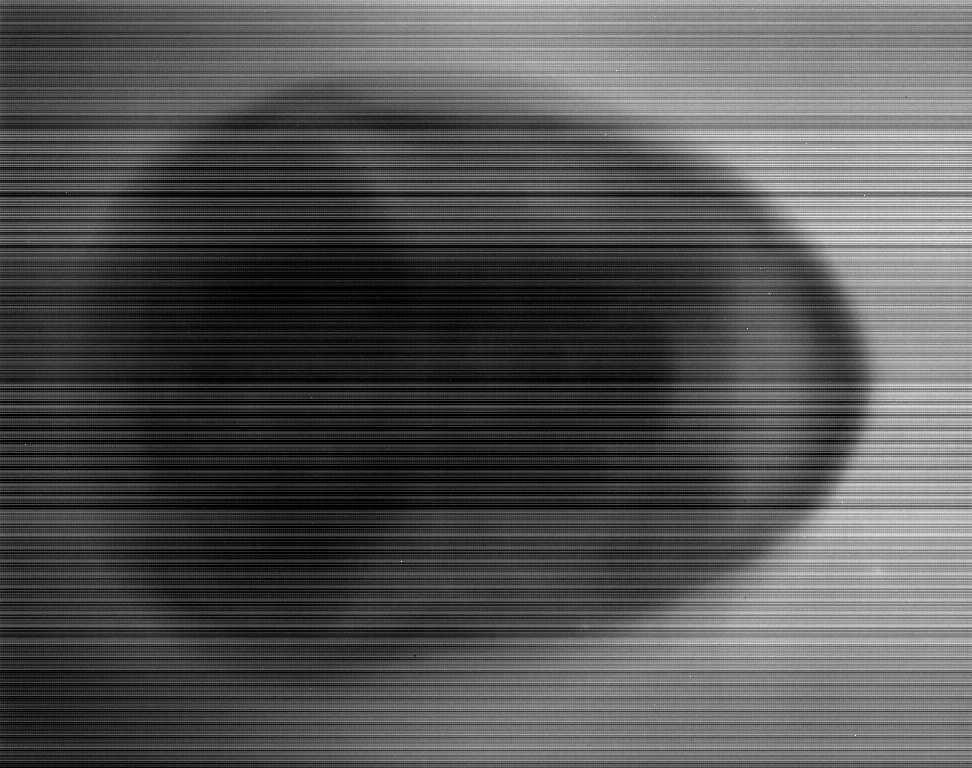}
                \subcaption*{[330,460]}
            \end{subfigure}
            \begin{subfigure}[c]{\imw\textwidth}
                \includegraphics[angle=90,width=\textwidth]{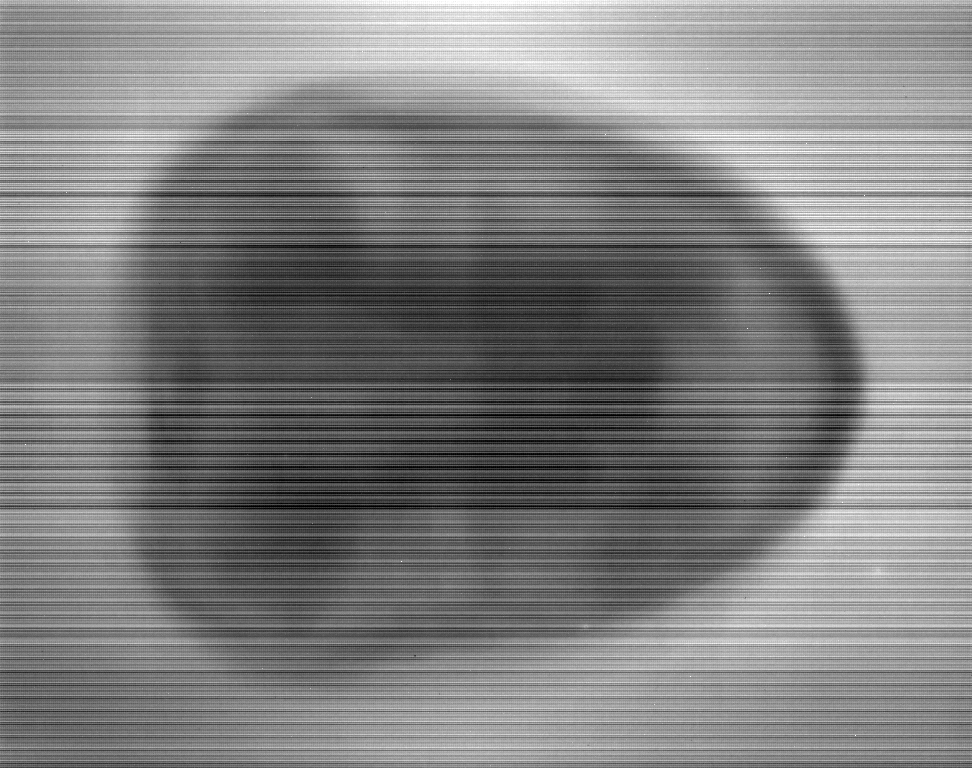}
                \subcaption*{[345,467]}
            \end{subfigure}
        \end{subfigure}
        \begin{subfigure}[b]{\textwidth}
            \begin{subfigure}[c]{0.05\textwidth}
                \caption*{\rotatebox{90}{pre-processed projections}}
            \end{subfigure}
            \begin{subfigure}[c]{\imw\textwidth}
                \includegraphics[angle=90,width=\textwidth]{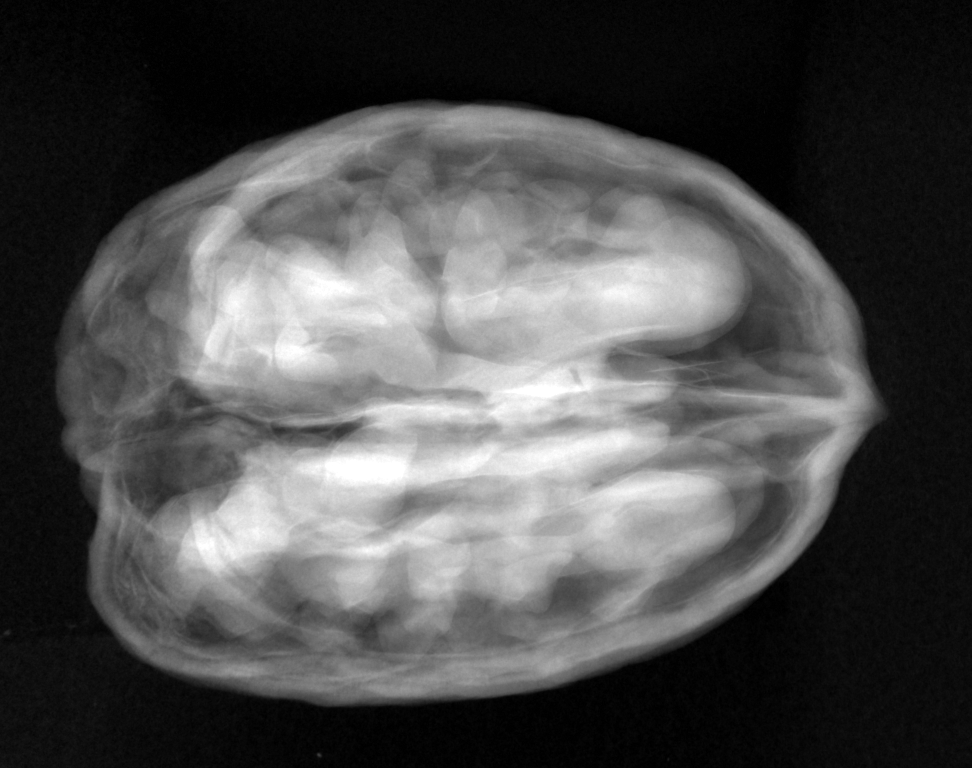}
                \subcaption*{[0,1]}
            \end{subfigure}
            \begin{subfigure}[c]{\imw\textwidth}
            \includegraphics[angle=90,width=\textwidth]{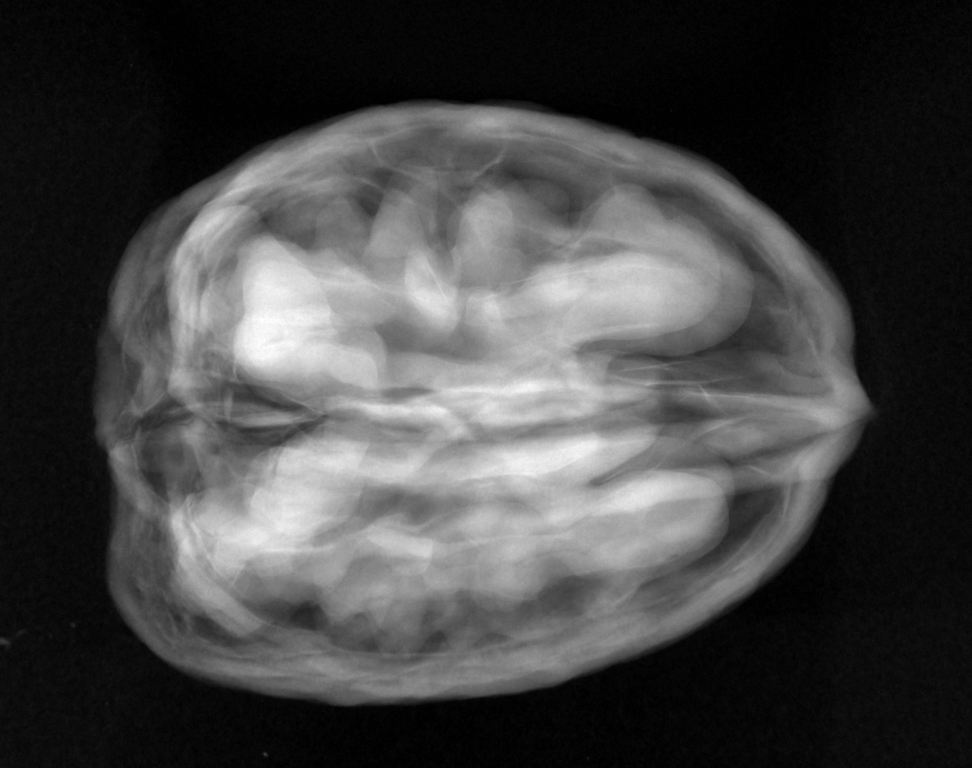}
            \subcaption*{[0,1]}
            \end{subfigure}
            \begin{subfigure}[c]{\imw\textwidth}
            \includegraphics[angle=90,width=\textwidth]{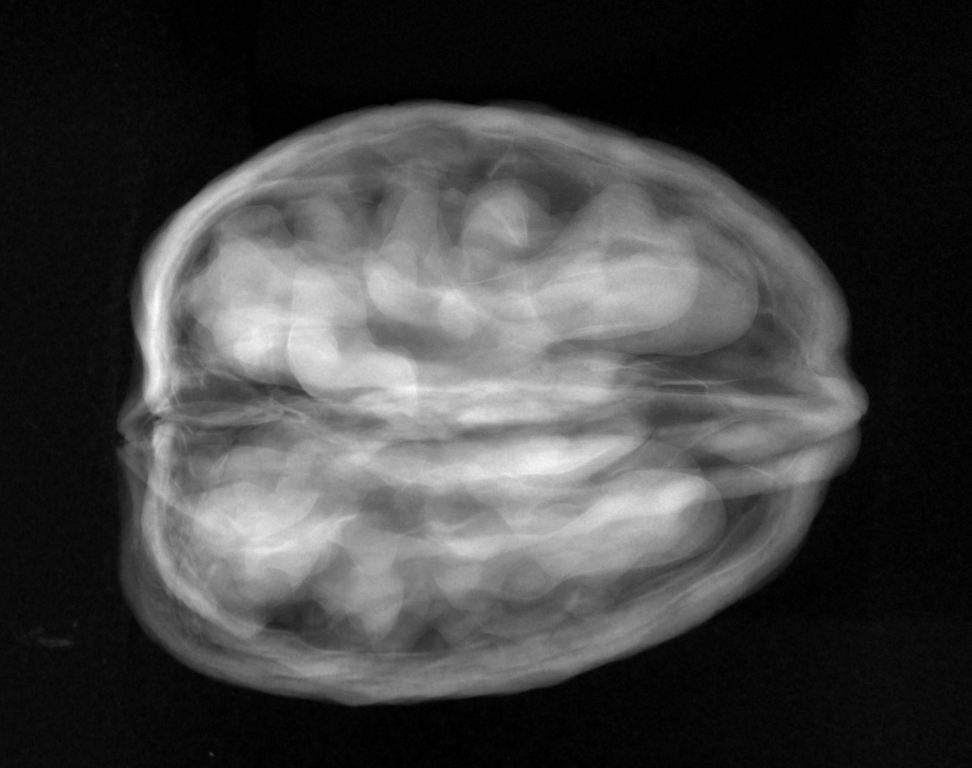}
            \subcaption*{[0,1]}
            \end{subfigure}
        \end{subfigure}
    \caption{Examples of the collected projections. From left to right, the position of the source varies. The dynamic range is indicated below each image.}
    \label{fig.projex}
\end{figure}

\subsection{Reconstructed Volumes}

\begin{figure}[tb]
    \centering
    \begin{subfigure}[b]{0.32\textwidth}
    \includegraphics{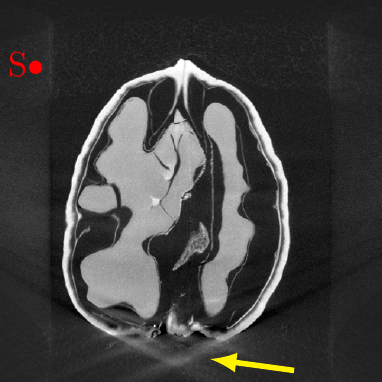}
    \end{subfigure}
    \begin{subfigure}[b]{0.32\textwidth}
        \includegraphics[]{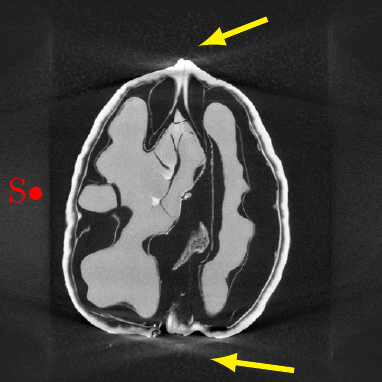}
    \end{subfigure}
    \begin{subfigure}[b]{0.32\textwidth}
        \includegraphics[]{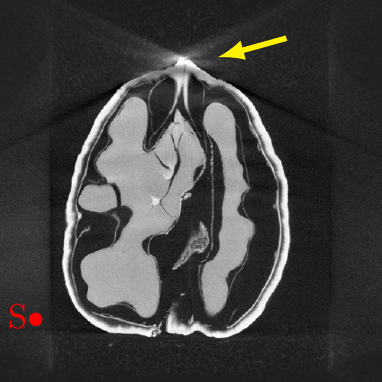}
    \end{subfigure}
    \\
    \vspace{0.08cm}
    \begin{subfigure}[b]{0.32\textwidth}
        \includegraphics[]{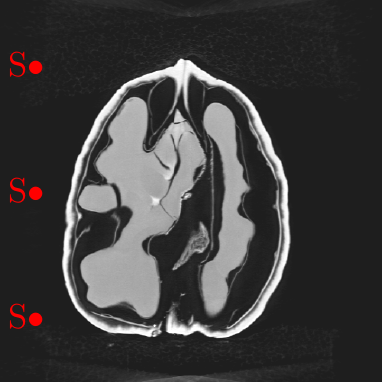}
    \end{subfigure}
    \caption{Vertical slice through reconstructed volumes from a single sample. Red dots indicate the source height for the circular orbit used for the reconstruction. Top row: FDK reconstruction from top, middle, and low source position. Yellow arrows point at the high cone angle artefacts. Bottom row: Iterative reconstruction from combined measurements.}
    \label{fig.reconall}
\end{figure}

Each projection image $P$ consist of raw photon counts per detector pixel that are distorted by off-set counts ("dark currents") and pixel-dependent sensitivities. Using the corresponding recorded dark-field image $D$ and flat-field image $F$, $P$ can be corrected and converted into a beam intensity loss image $I$ following the Beer-Lambert law as 
\begin{equation}
    I = - \log \left( \frac{P - D}{F - D} \right) \qquad .
\end{equation}
For each sample and each of the three source positions, a reconstruction was computed using the FDK algorithm~\cite{feldkamp1984} implemented in the ASTRA toolbox~\cite{vanaarle2016}. Then, the data from all source positions was combined to compute a high-quality reconstruction. This was done by solving a non-negativity constrained least-squares problem using 50 iterations of accelerated gradient descent~\cite{chambolle2016}. The corresponding forward and backward projection operators were implemented using the CUDA kernels in the ASTRA toolbox. In both cases, we chose a volume of $501^3$ voxels of size $100\mu\mathrm{m}^3$. The computation for one FDK with the data from one orbit took about $24\mathrm{s}$ on an NVIDIA GeForce GTX 1070, the iterative reconstruction of the complete data $56\mathrm{min}$. An example of the reconstructed volumes is shown in Figure~\ref{fig.reconall}: In the FDK reconstructions from single orbit data, the image regions with low beam incident angles are reconstructed well while strong artefacts can be seen in regions with high beam angle. They are caused by a combination of two factors: First, the circular orbit associated with a cone shaped beam does not fulfill Tuy's condition~\cite{tuy1983},  resulting in missing data in the measurement domain located around the rotation axis. Second, the FDK algorithm approximates the incoming beam by a collection of tilted fan-beams for each row of the flat detector. In contrast, the iterative reconstruction from the combined data is both sharp and artefact-free and can therefore be regarded as a ground truth reconstruction.

\subsection{Code Availability}

Python and MATLAB scripts for loading, pre-processing and reconstructing the projection data in the way described above are published on github\footnote{\href{https://github.com/cicwi/WalnutReconstructionCodes}{https://github.com/cicwi/WalnutReconstructionCodes}}. They make use of the ASTRA toolbox, which is openly available on \url{www.astra-toolbox.com} or accessible as a conda package\footnote{use \texttt{conda install -c astra-toolbox/label/dev astra-toolbox} to install the development version}. ASTRA is currently only fully supported for Windows and Linux\footnote{Installing it on Mac OS is possible but very involved and version-dependent.}. For obtaining a comparable scaling of the image intensities between FDK and iterative reconstructions, it is required to use a development version of the ASTRA toolbox more recent than 1.9.0dev. For each dataset, a text file containing information about motor positions (source 3D position, detector position and detector orientation) is provided and used by the aforementioned Python/MATLAB scripts to set up the reconstruction geometry. All reference reconstructions provided have been computed with the Python scripts. Furthermore, while the scripts allow to sub-sample the projections and to choose a different image resolution, the reference reconstructions were computed with all projections and within a volume of $501^3$ voxels of size $100\mu\mathrm{m}^3$ as mentioned above.


\section{Data Records}

The complete projection data for a single walnut and the corresponding reference reconstructions are shared as a single ZIP archive (ca. $6\mathrm{GB}$ per file). The $42$ resulting ZIP files (named \verb$Walnut1.zip - Walnut42.zip$, ca. $254.2\mathrm{GB}$ in total), were   
uploaded on zenodo\footnote{\href{https://zenodo.org}{https://zenodo.org}}, and had to be split up into several bundles to with separate DOIs: Samples 1-8 \cite{dataCit1}, samples 9-16 \cite{dataCit2}, samples 17-24 \cite{dataCit3}, samples 25-32 \cite{dataCit4}, samples 33-37 \cite{dataCit5} and samples 38-42 \cite{dataCit6}. Note, however, that each ZIP file can be downloaded separately via zenodo's web interface. \\
The ZIP file for the i$^{th}$ sample, \verb$Walnut<i>.zip$, contains a folder \verb$Walnut<i>/$ with the sub-folders \verb$Projections/$ and \verb$Reconstructions/$:
\begin{itemize}
    \item \verb!Projections/tubeV<j>/! contains the measured projection data with the source at position \verb!j!, where \verb!j=1/2/3! corresponds to the high/middle/low source position (cf. Figure~\ref{fig.trajectories}). Each of these folders contains the files:
    \begin{itemize}
        \item \verb!di000000.tif! is a TIFF file containing the dark-field measurement (cf. Figure~\ref{fig.projex}).
        \item \verb!io000000.tif! and \verb!io000001.tif! are TIFF files containing the flat-field measurements before and after the orbit was scanned (cf. Figure~\ref{fig.projex}).
        \item \verb!scan_<k>.tif! is a TIFF file containing the projection measurement at angle \verb!k! (cf. Figure~\ref{fig.projex}).
        \item \verb!scan_geom_original.geom! and \verb!scan_geom_corrected.geom! are text files describing the acquisition geometry of each angular projection. Their format and usage is explained in more detail in the following sections.
        \item \verb!data settings XRE.txt! and \verb!scan settings.txt! are text files automatically generated by the FleX-ray scanning software containing scan settings such as motor positions, source power or camera exposure time. We included them for completeness. 
        \item \verb!script_executed.txt! is a text file automatically generated by the FleX-ray scanning software containing a copy of the script executed by the scanner. We included it for completeness.
    \end{itemize}
    \item \verb$Reconstructions$ contains the reference reconstruction as described above, stored as TIFF files each containing a single $x$-slice of the volume:
    \begin{itemize}
        \item \verb!fdk_pos<j>_<k>! contains the \verb!k!$^{th}$ $x$-slice of the FDK reconstruction computed from the projection data acquired at source position \verb!j! (cf. Figure~\ref{fig.reconall}).
        \item \verb!full_AGD_50_<k>! contains the \verb!k!$^{th}$ $x$-slice of the ground truth reconstruction computed by $50$ iterations of accelerated gradient descent (cf. Figure~\ref{fig.reconall}).
    \end{itemize}{}
\end{itemize}


\section{Technical Validation}

The FleX-ray scanner is subject to regular maintenance and calibration. Furthermore, a visual inspection of all projections for each sample was carried out to ensure that the collected data does not suffer from over-saturation and the sample was always in the field of view. The reconstructed volumes were inspected to ensure that the correction of geometric  distortions such as in-plane rotation tilt of the detector was successful. 
For the iterative reconstruction from the combined data (ground truth reconstruction), the registration of the scanning geometries from the single orbits had to be corrected manually due to mechanical inaccuracies in the motors positions reported by the scanner. For this, three volumes corresponding to the three orbits were reconstructed first and then manually co-registered using rigid transformations. Corresponding corrected geometry description text files that are used in the combined reconstruction are provided (\verb!scan_geom_corrected.geom!). Samples for which this procedure did not succeed were discarded. For completeness, the original geometry description text files as deduced from the reported motor positions are also provided (\verb!scan_geom_original.geom!).

\section{Usage Notes}

\subsection{Projection Data}

\begin{figure}
    \centering
    \includegraphics[]{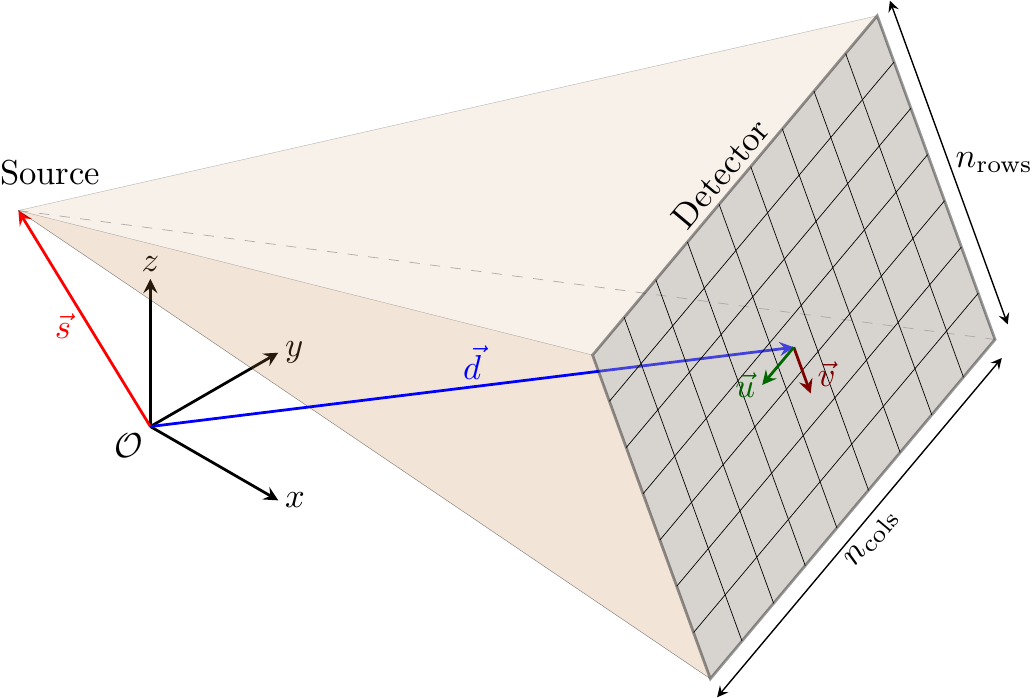}
    \caption{\label{fig.geom}Parametrization of the cone-beam geometry: Each projection is described by $(s_x, s_y, s_z, d_x, d_y, d_z, u_x, u_y, u_z, v_x, v_y, v_z)$}
    \end{figure}

The projection data for each sample is shared as a collection of 16 bit unsigned integer TIFF files containing the raw photon counts per detector pixel. They can be interpreted and manipulated by most common image visualization software such as ImageJ~\cite{imagej} or scientific computing languages such as MATLAB or Python, e.g., through the matplotlib module for the latter. In order to be used by most tomographic reconstruction algorithms, they need to be pre-processed as described above and exemplified in the provided scripts. Each row of the geometry description files (\verb!scan_geom_*.geom!) describes the geometry of one of the acquired projections by 12 floating point numbers: source $x$ position, source $y$ position, source $z$ position, detector center $x$ position, detector center $z$ position, detector center $z$ position, detector 3D basis vector from pixel $(0,0)$ to pixel $(1,0)$, and detector 3D basis vector from pixel $(0,0)$ to pixel $(0,1)$. This parametrization is illustrated in Figure~\ref{fig.geom}.

\subsection{Reconstructed Volumes}

In principle, the four reconstructions described in the previous sections (cf. Figure~\ref{fig.reconall}) can be computed from the projection data with the scripts provided. Depending on the available computational resources this may, however, require a lot of computing time. For this reason, we share the reconstructions, too. They can also be used as a comparison to test novel reconstruction algorithms, or as an image collection for image analysis tasks. Each volume is released as a collection of 32 bit floating point TIFF files, where every single file is one axial slice through the volume as described above. As for the projection data, open source software is available for visualization and manipulation of such files.

\subsection{Further Usage}

The reconstruction scripts can easily be modified to generate different kind of artefacts and tackle different problems related to tomographic reconstruction. To create a limited or sparse-angle (low-dose) problems, one can simply load subsets of the projection data. To mimic a super-resolution experiment, the projection data can be artificially binned into larger pixels. In every case, the iterative reconstruction from the full data set can be used as a ground truth.



\section*{Acknowledgements}

This work was supported by the Netherlands Organisation for Scientific Research (NWO 613.009.106, 639.073.506). 
The authors would like to thank Alexander Kostenko for his help in using the FleX-ray Lab and sample preparations, and Nicola Vigan\`{o} for his help with the image registration.

\section*{Author Contributions}

HDS, FL, MvE and KJB conceptualized the study and designed the experiment. 
HDS, GC and SBC set up the experiment and performed the data acquisition.
HDS performed the data processing, inspection and geometry correction.
HDS and FL wrote the reconstruction scripts and the the main parts of the manuscript. 
All authors contributed to the discussion and finalization of the manuscript and approved it.




\section*{Competing Financial Interests}

The authors declare no competing financial interests.

\end{document}